\documentclass[preprint,12pt]{elsarticle}
\usepackage{float}
\usepackage[parfill]{parskip}    
\usepackage{graphicx}
\usepackage{amssymb}
\usepackage{epstopdf}
\usepackage{color}
\usepackage{float}
\usepackage{amsmath}
\usepackage{orcidlink}
\usepackage{latexsym}
\usepackage{amsmath}
\usepackage{amssymb}
\usepackage{amsfonts}
\usepackage{blindtext}
\usepackage{subfigure}

\begin{document}
\title{Cosmological FLRW phase transitions under exponential corrected entropy }

\author{Rodrigo Rivadeneira-Caro}
\ead{rodrigo.rivadeneira.c@mail.pucv.cl}

\address{Facultad de Física, Pontificia Universidad Católica de Chile, Avenida Vicuña Mackena 4860, Santiago, Chile}

\address{
Instituto de Física, Pontificia Universidad Católica de Valparaíso, Casilla 4950, Valparaíso, Chile.}

\author{Joel F. Saavedra \orcidlink{0000-0002-1430-3008}}
\ead{joel.saavedra@pucv.cl}
\address{
Instituto de Física, Pontificia Universidad Católica de Valparaíso, Casilla 4950, Valparaíso, Chile.}

\author{Francisco Tello-Ortiz \orcidlink{0000-0002-7104-5746}}
\ead{francisco.tello@ufrontera.cl}
\address{Departamento de Ciencias Físicas, Universidad de La Frontera, Casilla 54-D, 4811186 Temuco, Chile.}

\begin{abstract}

This work considers how exponential corrections to the Bekenstein-Hawking entropy formula affect the thermodynamic behavior of the FLRW cosmological model. These corrections significantly alter the form of the Friedman field equations, leading to nontrivial phase transition behavior. For negative values of the tracking parameter $\alpha$, the system presents first-order phase transitions above the critical temperature, and for positive $\alpha$, the system undergoes a reentrant phase transition. As these corrections are presumably relevant at the early Universe stage, to corroborate the presence of some potential vestige of this contribution in the current era, a study has been carried out comparing observational data and current values of the Hubble parameter.   

\end{abstract}

\maketitle

\section{Introduction}
It is well-known from the pioneering works \cite{Bekenstein:1973ur,Bardeen:1973gs,Hawking:1975vcx,Gibbons:1976ue}, that a black hole (BH) has an entropy proportional to its area defined by its event horizon: $S=A/4$ with $A=4\pi r^{2}_{H}$ and $r_{H}$ the BH event horizon. In general, this result applies to all systems in Einstein's theory; however, it is necessary to consider the nature of the model. In this concern, for a cosmological manifold, for example, a Friedmann-Robertson-Walker (FRW) metric, the entropy remains $S=A/4$, but this time the area is given by $A=4\pi r^{2}_{AH}$, where $AH$ stands for the apparent horizon in this case. If one changes the theory, the entropy adopts a different form, generally the so-called Bekenstein-Hawking entropy \cite{Bekenstein:1973ur,Hawking:1975vcx} plus some corrections. The primary difference in adopting various solutions lies in identifying the relevant horizon. This result can be expected, as it is derived from the fundamental aspects underlying the theory \cite{Gibbons:1976ue}. Another universal result is related to the intrinsic connection between gravity and thermodynamics, where Einstein's field equations are interpreted as the thermodynamic law and vice versa \cite{Jacobson:1995ab}. In the context of cosmological scenarios, since the system is naturally a dynamical system, the thermodynamic laws should be reformulated \cite{Kodama:1979vn,Hayward:1993wb,Hayward:1994bu,Hayward:1997jp}.

The deep connection between thermodynamics and gravity enables us to relate local energy-momentum variables to the relevant thermodynamic potentials. In simple words, given the geometric interpretation of temperature through surface gravity, the gravitational equations of motion are easy to obtain. This technique has been widely used in several gravity theories, such as Lovelock and Gauss-Bonnet gravity \cite{Cai:2005ra}, $f(R)$ theories \cite{Cai:2006pa}, brane-world \cite{Cai:2006rs}, massive gravity \cite{Li:2013fop}, and other modified gravity theories \cite{Sebastiani:2023brr,Nojiri:2025gkq,Nojiri:2024zdu,Nojiri:2023wzz,Nojiri:2022nmu,Bamba:2018zil}. On the other hand, if one supplements the system with the entropy expression, in principle, any field equations can be obtained, accounting for the correction introduced for the entropy form. In this context, several entropy proposals have been translated into the gravitational context. For example, the relativistic Kaniadakis entropy \cite{Kaniadakis:2002zz,Kaniadakis:2005zk}, loop quantum gravity entropy \cite{Zhang:2008gt}, and Barrow's entropy \cite{Barrow:2020tzx}, to name a few, have been used in the cosmological setting to reveal how modifications come into Friedman equations\footnote{It is worth mentioning that all these entropies are particular cases of a generalized one put forward in \cite{Nojiri:2022aof,Nojiri:2022dkr} } \cite{Sheykhi:2010zz,Sheykhi:2021fwh,Sheykhi:2023aqa}. Interestingly, corrections to the entropy–area law are usually logarithmic, power-law, or non-extensive. All these effects modify the Friedmann equations in distinct ways, producing new dynamical behavior in the cosmic expansion and leaving potential imprints on observable quantities.

A natural question arising from these developments is whether horizon thermodynamics can reveal features of cosmological evolution that are inaccessible within the standard area law. Modified entropies are known to influence gravitational dynamics across diverse frameworks. Yet, their effects in FLRW scenarios remain only partially understood, particularly regarding the emergence of critical behavior and changes in stability at the level of the $AH$. Since the horizon in a dynamical spacetime acts as a thermodynamic boundary with a well-defined temperature and energy flux, even subleading corrections to the entropy may reorganize its phase structure and, consequently, alter the effective evolution equations governing the expansion. This motivates a systematic examination of whether specific entropy deformations can imprint distinguishable thermodynamic signatures—such as heat capacity anomalies, critical points, or non-standard phase transitions—while remaining consistent with the empirical success of $\Lambda$-CDM. Exploring this question provides not only a conceptual probe of horizon thermodynamics but also a potential pathway to identify subtle deviations in the cosmic expansion history that may originate from quantum-gravitational microstructure.
In this broader context, we use exponential entropy corrections. Motivated by recent developments in the counting of quantum gravitational microstates \cite{Chatterjee:2020iuf}, the entropy receives an additive term of the form
\begin{equation}
S = \frac{A}{4\ell_{P}^{2}} +  \, e^{-A\delta/4\ell_{P}^{2}}.
\end{equation}
This correction decays exponentially with area, ensuring that for large horizons the standard Bekenstein–Hawking law is recovered, while at small scales-such as those relevant for the early Universe—the modification can dominate. Exponential entropy, therefore, provides a natural mechanism for probing quantum effects during the earliest stages of cosmic evolution.

From a thermodynamic perspective, these modifications allow us to study novel phase structures. The $AH$, rather than being a passive boundary, may display critical points, stability changes, and first-order phase transitions analogous to those found in BH thermodynamics. So, these corrections at hand raise the possibility that the thermodynamic history of the Universe may itself involve transitions between distinct phases, governed by horizon thermodynamics.

{In this regard, it is worth emphasizing that the extension of entropic corrections beyond the black hole scenario into cosmology has been explored in several works \cite{Jacobson:1995ab, Cai:2005ra, Padmanabhan2005,AkbarCai2007,Hayward2006}. The fact that the exponentially corrected entropy does not vanish in the limit $A \to 0$ should not be seen as an inconsistency, but rather as an indication that the effective description ceases to be valid in the very early Universe, where quantum-gravitational effects are expected to dominate. Indeed, other recent proposals, such as the Tsallis and Kaniadakis entropies \cite{Kaniadakis:2002zz,Tsallis1988,Lymperis2022,Alruwaili:2025bwf,Jawad:2019vqa}, or the fractal entropy of Barrow \cite{Barrow:2020tzx,Jawad:2025uqu}, also exhibit non-trivial behavior in limiting regimes, suggesting that the strict monotonicity of $S(A)$ might be relaxed in a holographic quantum framework. Moreover, the existence of a minimum in the entropic function can be interpreted as the statistical origin of critical phenomena in cosmology, analogous to phase transitions in thermodynamic systems \cite{KubiznakMann2012, WeiLiu2020}. Within this perspective, the phase transition observed using these corrected entropies should not be regarded as a mere mathematical artifact, but rather as a possible manifestation of a change in physical regime during cosmic evolution, supporting the idea that entropic corrections beyond the Bekenstein–Hawking entropy may capture essential features of the Universe's dynamics.

Following this direction, the present work is devoted to studying an FLRW cosmological model with exponential entropy corrections. In Sect. \ref{sec2} we first revisit the derivation of the Friedmann equations from the unified first law of thermodynamics, now modified by the exponential entropy. In Section \ref{sec3}, we construct the effective equation of state for horizon thermodynamics, paying particular attention to the role of the  tracking $\alpha$ controlling the exponential entropy corrections. The analysis of the heat capacity (Sect. \ref{sec3}) and Gibbs free energy in Sect. \ref{sec4}, reveals the conditions under which the system experiences phase transitions, critical points, or instabilities. These thermodynamic insights are subsequently connected to cosmological dynamics in 
Sect. \ref{sec5}, where we examine how the modified equations alter the Hubble expansion history $H(z)$ and compare the predictions against observational probes such as cosmic chronometers and the Pantheon+ supernova dataset. Finally, in Sect. \ref{sec6} we emphasize the potential relevance of exponential entropy corrections to outstanding cosmological problems, including the current $H_{0}$ tension, and we summarize the broader implications of a universe whose horizon exhibits a rich thermodynamic phase structure.

\section{Friedmann equations revised}\label{sec2} 

It was shown in \cite{Chatterjee:2020iuf} that counting those quantum micro-states residing on the BH event horizon leads to an exponential correction to the Bekenstein-Hawking entropy formula. Specifically, the corrected entropy reads\footnote{When quantum corrections are taken into account, the full entropy is \begin{equation}
S=\frac{{A}}{4 \ell_P^2}+\gamma \ln \frac{{A}}{4 \ell_P^2}+\beta \frac{4 \ell_P^2}{{A}}+\cdots+\exp \left(-\delta \frac{{A}}{4 \ell_P^2}\right)+\cdots
\end{equation}
where $\gamma$ and $\beta$ and are dimensionless constants or order of unity \cite{jzhang}.
It should be pointed out that logarithmic corrections do not always arise \cite{Ghosh:2012jf}.}
\begin{equation}\label{entropy}
S=\frac{A}{4\ell_P^2}+  e^{-A \delta / 4\ell_P^2},
\end{equation}
being $\ell_{p}$ Planck's length, $\delta$ a universal dimensionless constant of order of unity and $A$ the BH horizon area. However, it is possible to re-express the entropy in the following way \cite{Okcu:2024llu}
\begin{equation}\label{eq3}
S=\frac{A}{4}+\alpha e^{-A/4},    
\end{equation}
where $\alpha$ is a  dimensionless constant, the so-called tracking parameter\footnote{This is because when Planckian's units are used, that is, when $G=c=\hbar=k_{\text{B}}=1$, the entropy $S$ becomes dimensionless, and for the sake of simplicity we have taken $\delta=1$.}.

In the cosmological scenario, the mentioned area corresponds to the $AH$ area.  An essential feature of the corrected entropy (\ref{entropy}) is that, for large areas, the exponential correction is negligible. Nevertheless, within the cosmological framework, depending on the epoch of the Universe's evolution, this can make a substantial contribution.  

To conduct the thermodynamic study, it is necessary to first define all the physical quantities involved. Given that such a study is carried out on a particular surface in a gravitational context (i.e., the $AH$ in cosmological framework and the event horizon for the BH case), it is necessary to introduce the geometric model from which such a surface is derived. In this case, the FRW cosmological model given by
\begin{equation}\label{flrw}
    ds^{2}=-dt^{2}+a^{2}(t)\left(\frac{dr^{2}}{1-kr^{2}}+r^{2}d\Omega^{2}\right).
\end{equation}

In general, to deal with the thermodynamics of $AH$s, one needs to modify the usual approach based on the  thermodynamics first law, employed in the static/stationary BHs case \cite{ Bekenstein:1973ur,Bardeen:1973gs, Hawking:1975vcx, Gibbons:1976ue}. In this regard, Hayward developed this approach in a series of articles \cite{Hayward:1993wb, Hayward:1994bu, Hayward:1997jp}. The main ingredients of this formulation are: i) the Misner--Sharp (MS) mass $M_{AH}$, interpreted as the internal energy, ii) the  Hayward-Kodama (HK) surface gravity (related with the temperature of the $AH$) and iii) the areal volume $V_{AH}=\frac{4\pi}{3}R^{3}_{AH}$. Additionally, it is necessary to introduce new quantities to reach the desired thermodynamic description. These quantities are: iv) the  work density $W$ and the v) energy--flux $\psi_{i}$ across the $AH$
\cite{Hayward:1997jp}. The definition of these new pieces is given by
\begin{equation}\label{workdensity}
    W\equiv-\frac{1}{2}h_{ij}T^{ij},
\end{equation}
and
\begin{equation}
    \psi_{i}\equiv T^{j}_{i}\nabla_{i}R+W\nabla_{i}R,
\end{equation}
respectively. Furthermore, $T^{ij}$ represents the projected energy-momentum tensor components.
Putting together all these objects, the thermodynamic unified first law (UFL) acquires the following form \cite{Hayward:1993wb, Hayward:1994bu, Hayward:1997jp}
\begin{equation}
    \nabla_{i}M=A\psi_{i}+W\nabla_{i}V,
\end{equation}
or by identifying the mass $M$ with the internal energy $E$ one obtains 
\begin{equation}\label{UFL}
    \nabla_{i}E=A\psi_{i}+W\nabla_{i}V.
\end{equation}
It is worth mentioning that the term $A\psi_{i}$ is called the energy-supply vector.

  Now, to determine the $AH$, the line element (\ref{flrw}) is written as a warped product between a two-dimensional manifold $\mathcal{M}_{2}$ (the $t-r$ plane) and a two-sphere $\mathbb{S}^{2}$
as follows
\begin{equation}
    ds^{2}=h_{ij}dx^{i}dx^{j} + R^{2}d\Omega^{2}.
\end{equation}
Here $h_{ij}$ is the induced metric on the variety $\mathcal{M}_{2}$ and $R(t,r)\equiv a(t)r$ is the physical radius. So, it is not difficult to check that the $AH$ is the solution of the differential equation \cite{Faraoni:2011hf,Faraoni:2015ula}
\begin{equation}\label{AHsolu}
    h^{ij}\nabla_i R \nabla_j R = 0
\Rightarrow
    R_{\text{AH}}=\frac{1}{H},
\end{equation}
where $H\equiv \dot{a}/a$ is the Hubble constant and we have assumed, without generality, a spatially flat Universe, that is, $k=0$. In this way, the radius of the $AH$ coincides with the cosmological horizon. Here, the Latin indices run over $i, j, t,r$. Next, the trace of the energy-momentum tensor $h_{ij}T^{ij}$ on the $t-r$ plane orthogonal to the two-spheres is given by $h_{ij}T^{ij}=p-\rho$, being $p$ the isotropic pressure and $\rho$ the density of the perfect fluid filling the Universe.

The projection of the UFL (\ref{UFL}) along the $AH$ yields \citep{Hayward:1997jp} 
\begin{equation}\label{projected}  z^{i}\nabla_{i}E=\frac{\kappa_{\text{HK}}}{8\pi}z^{i}\nabla_{i}A+Wz^{i}\nabla_{i}V,
\end{equation}
identifying $A\psi_{i}=\frac{\kappa_{\text{HK}}}{8\pi}z^{i}\nabla_{i}A$  \citep{Hayward:1997jp,Cai:2006rs} and $z^{i}$ being a tangent vector to the $AH$. This allows us to identify the so-called HK surface gravity $\kappa_{\text{HK}}$. Considering that this is a geometric object (independent of the underlying theory), it is defined as \citep{Hayward:1997jp}
\begin{equation}\label{SG}
  \kappa_{\text{HK}}=\frac{1}{2\sqrt{-h}}\partial_{i}\left(\sqrt{-h}h^{ij}\partial_{j}R\right),  
\end{equation}
where $h\equiv \text{det}(h_{ij})$. The connection between the surface gravity and temperature $T$ is  
\begin{equation}\label{eq12}
    T=\frac{ \kappa_{\text{HK}}}{2\pi}.
\end{equation}

After replacing the surface gravity with the temperature in the projected UFL (\ref{projected}), the first term on the right-hand side can be recognized as the Clausius relation, $dE=-TdS$, where, by using (\ref{eq3}) and some algebraic reduction, one arrives at 
\begin{equation}\label{eq13}
-\frac{4 \pi}{3} \rho=-\frac{1}{2 R_{AH}^2}+\frac{\alpha}{2 R_{AH}^2}\left[e^{-\pi R_{AH}^2}+\pi R_{AH}^2 \text{Ei}\left(-\pi R_{AH}^2\right)\right],
\end{equation}
and
\begin{equation}\label{eq14}
-4 \pi(\rho+p)=\left(1-\alpha e^{-\pi R_{AH}^2}\right) \dot{H},
\end{equation}
where (\ref{AHsolu}) has been used. Here, Ei(x) is the exponential integral. It is worth mentioning that these equations were obtained for the first time in \cite{Okcu:2024llu} in a slightly different manner.  

Now that we know the explicit expressions for $\rho$ and $p$, it is possible to obtain a general equation of state (EoS) for describing the thermodynamics of the system.

\section{The equation of state}\label{sec3}

At this level, the EoS of the system is recognized as the work density (\ref{workdensity}), that is,
\begin{equation}
P(R_{AH},T)\equiv W=\frac{1}{2}\left(\rho-p\right),   \end{equation}
where the radius of the $AH$ is connected with the 
the specific volume as follows $v=2\ell^{2}_{P}R_{AH}=2R_{AH}$ \citep{Kubiznak:2012wp}. Providing, in this way, a relation of the type
\begin{equation}
    P=P(v,T),
\end{equation}
as usual. However, before obtaining the desired EoS, it is pertinent to discuss the global sign of the temperature. The Eq. (\ref{eq12}) gives the usual relation between surface gravity and temperature. So, at first sight, the final global sign of the temperature depends on the final sign of surface gravity. This means that, in principle, both positive and negative temperatures are allowed. This is because the surface gravity is not always positive-definite on $AH$. This fact is intimately related to the inner/outer feature of the $AH$. For an inner $AH$, the surface gravity is negative in nature, since the congruence along the ingoing null direction vanishes on the $AH$, that is, $\theta_{-}=0$ and the congruence along outgoing null direction is positive, that is, $\theta_{+}>0$ (the future case stands for the opposite situation: $\theta_{+}=0$ and $\theta_{-}<0$). Besides, the past/future characteristic is related with the Lie derivative of the ingoing null congruence along the null outgoing direction, that is, $\mathcal{L}_{+}\theta_{-}>0$ for past and     $\mathcal{L}_{+}\theta_{-}<0$ for future. Therefore, one has the following four cases: i) inner-past $AH$ \{$\theta_{-}=0$; $\theta_{+}>0$; $\mathcal{L}_{+}\theta_{-}>0$\}, ii) outer-past \{$\theta_{-}=0$; $\theta_{+}>0$; $\mathcal{L}_{+}\theta_{-}<0$\}, iii) inner-future $AH$ \{$\theta_{+}=0$; $\theta_{-}<0$; $\mathcal{L}_{-}\theta_{+}>0$\} and outer-future $AH$ \{$\theta_{+}=0$; $\theta_{-}<0$; $\mathcal{L}_{-}\theta_{+}<0$\} \cite{Helou:2015yqa,Helou:2015zma}. On the other hand, the specific matter content dominating the different epochs of the Universe's evolution corresponds to inner $AH$. In particular, inflation, dust, and dark energy correspond to a past-inner\footnote{In the pure GR case, radiation leads to a degenerated $AH$, that is, an $AH$ with vanishing surface gravity.} $AH$. Moreover, in the hypothetical case where a stiff matter content drove the initial Universe, the $AH$ corresponds to an outer-past one. Therefore, in any case, the $AH$ in the Universe's history is past in nature. This is an important fact, since $\mathcal{L}_{+}\theta_{-} \propto-\kappa_{\text{HK}}$. Thus, if $\kappa_{\text{HK}}<0$, one has an inner-past $AH$; otherwise, it will be outer-past. In this way, the final sign of the temperature is determined by the inner/outer and past/future features of the $AH$. In this regard, an inner past $AH$ will have a positive temperature, while an inner future will have a negative one. 

Therefore, to further elucidate and clarify whether these modified Friedmann equations admit positive, negative temperatures, or both, it is important to express the surface gravity (\ref{SG}) evaluated at the $AH$ in terms of the thermodynamic variables $\rho$ and $p$. So, using Eqs. (\ref{eq13})-(\ref{eq14}) one gets
\begin{equation}
\kappa_{\text{HK}}\bigg{|}_{AH}=-H+\frac{2 \pi(1+\omega) \rho}{H\left(1-\alpha e^{-\pi / H^2}\right)},
\end{equation}
where as usual, the following local barotropic EoS has been used
\begin{equation}
    p=\omega\rho,
\end{equation}
being $\omega$ the equation of state parameter. As the exponential correction decreases with increasing area (see Eq. (\ref{eq3})), we are going to consider those scenarios corresponding to early Universe stage, that is, those phases where $\omega$ takes the following numerical values: i) $\omega=-1$ (inflation), ii) $\omega=1/3$ (radiation), iii) $\omega=0$ (matter) and iv) $\omega=1$ (stiff matter). It is worth mentioning that, from the theoretical point of view, the early evolution of the Universe could have been dominated by stiff matter distributions \cite{Banks:2008ep}. Additionally, one needs to consider the parameter $\ alpha$'s signature and magnitude. For this purpose, Tables \ref{table1} and \ref{table2} summarize the cases where the surface gravity can be positive or negative, depending on the matter content and $\alpha$ signature. Remembering that for an FLRW metric, the $AH$ is always a past $AH$. So, $T\propto -\kappa_{\text{HK}}$. It is observed in Table \ref{table1} that for both $\alpha<0$ and $\alpha>0$, it is not possible to obtain the Universe facing the inflation epoch. On the other hand, in table \ref{table2}, all early eras are allowed independently of the $\alpha$ sign. $\alpha$ plays a major role in determining part of the causal structure of the $AH$, where it is possible to have for a FLRW cosmology an inner-past $AH$ and an outer-past $AH$. This will determine the final sign of the temperature. 

Now, before to conclude whether the model admits positive or negative temperature, it is relevant to discuss about $\alpha$ magnitude to further support the information provided in tables \ref{table1} and \ref{table2}. To do this, we are going to analyze the satisfaction of a mandatory requirement for any entropy, that is, $dS/dt\geq 0$. From Eq. (\ref{eq3}) one gets 
\begin{equation}
\frac{dS}{dt}=\frac{dS}{dA}\frac{dA}{dt} = \left(\frac{1}{4} - \frac{\alpha}{4} e^{-A/4}\right)\frac{dA}{dt}.
\end{equation}
Given that $\frac{dA}{dt}>0$, the Universe is expanding, the increasing entropy condition is satisfied if and only if $\alpha\leq e^{A/4}$. This implies that negative values are allowed for the tracking parameter.

Now, the sign of the effective surface gravity,
\begin{equation}
\kappa = -H + \frac{2\pi(1+\omega)\rho}{H(1 - \alpha e^{-\pi/H^2})},
\end{equation}
controls the tunneling temperature \(T = -\kappa / 2\pi\) and the causal character of the horizon (inner vs. outer). In the early universe (\(H^2 \gg 1\)), the exponential factor is moderate: \(e^{-\pi/H^2} \sim 0.7\). To invert the sign of \(\kappa\), it is necessary that the correction term
\(\alpha e^{-\pi/H^2}\) be of order unity. This implies that:
$\alpha = \mathcal{O}(1)$ is necessary to induce a thermal phase transition. In summary, the following conclusions emerge:
\begin{itemize}
    \item If \(\alpha > 0\), the entropy increases for small \(A\), and the correction can induce \(\kappa > 0\) (i.e., negative temperature) in radiation- or matter-dominated phases. This requires \(\alpha \gtrsim 0.1 - 1\), depending on \(\omega\).
    \item If \(\alpha < 0\), the correction suppresses entropy but robustly ensures \(\kappa < 0\) (positive temperature). 
\end{itemize}

Tables \ref{table1}, \ref{table2}, \ref{table3}, and \ref{table4} summarize all possibilities for the signature of the surface gravity, taking into account the signature of the $\alpha$ parameter and matter distributions present in the early Universe stage. 

\begin{table}[H]
\centering
\setlength{\tabcolsep}{1pt} 
\renewcommand{\arraystretch}{1.3} 
\begin{tabular}{|c|c|c|c|c|l|}
\hline
\textbf{Matter type} & \(\omega\) & \(1+\omega\) & \(\kappa > 0\) with \(\alpha > 0\) & \(\kappa > 0\) with \(\alpha < 0\) & \textbf{Comment} \\
\hline
Inflation & \(-1\) & \(0\) & \ No &  No & Positive term vanishes \\
\hline
Dust (Matter) & \(0\) & \(1\) & \ Possible &  Very unlikely & Requires large \(\rho\) \\
\hline
Radiation & \(1/3\) & \(4/3\) & \ Likely &  Unlikely & Holds at early times \\
\hline
Stiff matter & \(1\) & \(2\) & \ Very likely &  Only if \(\rho \gg 1\) & Strong positive contribution \\
\hline
\end{tabular}
\caption{Conditions under which \(\kappa > 0\) is possible, for different fluids and signs of \(\alpha\), in the early Universe (\(H^2 \gg 1\)).}
\label{table1}
\end{table}

\begin{table}[H]
\centering
\setlength{\tabcolsep}{0.5pt} 
\renewcommand{\arraystretch}{1.3} 
\begin{tabular}{|c|c|c|c|c|l|}
\hline
\textbf{Matter type} & \(\omega\) & \(1+\omega\) & \(\kappa < 0\) with \(\alpha > 0\) & \(\kappa < 0\) with \(\alpha < 0\) & \textbf{Comment} \\
\hline
Inflation & \(-1\) & \(0\) & \ Always &  Always & Positive term vanishes, \(\kappa = -H\) \\
\hline
Dust (Matter) & \(0\) & \(1\) & \ If \(\alpha \lesssim 0.2\) & Always & Large \(\alpha\) may flip sign \\
\hline
Radiation & \(1/3\) & \(4/3\) &  Only if \(\alpha \ll 0.1\) &  Always & \(\kappa\) flips sign if correction is strong \\
\hline
Stiff matter & \(1\) & \(2\) & \ No & \ Likely if \(\rho \gg 1\) & \(\kappa < 0\) difficult with \(\alpha > 0\) \\
\hline
\end{tabular}
\caption{Conditions under which \(\kappa < 0\) (i.e., \(T > 0\)) holds, depending on the fluid and the sign of \(\alpha\), in the early Universe (\(H^2 \gg 1\)).}
\label{table2}
\end{table}

\begin{table}[H]
\centering
\setlength{\tabcolsep}{10pt}
\renewcommand{\arraystretch}{1.3}
\begin{tabular}{|c|c|c|c|c|}
\hline
\textbf{Matter type} & \(\omega\) & \(\kappa\) sign & Temperature \(T = -\kappa / 2\pi\) & \textbf{Horizon type} \\
\hline
Inflation & \(-1\) & \(\kappa < 0\) & \(T > 0\) & Past–inner (radiative) \\
\hline
Dust (Matter) & \(0\) & \(\kappa > 0\) & \(T < 0\) & Past–outer (non-radiative) \\
\hline
Radiation & \(1/3\) & \(\kappa > 0\) & \(T < 0\) & Past–outer (non-radiative) \\
\hline
Stiff matter & \(1\) & \(\kappa > 0\) & \(T < 0\) & Past–outer (non-radiative) \\
\hline
\end{tabular}
\caption{Behavior of the surface gravity \(\kappa\), temperature, and horizon type for different cosmic fluids in the early universe (\(H^2 \gg 1\)) with exponential entropy correction parameter \(\alpha = 1\).}
\label{table3}
\end{table}

\begin{table}[H]
\centering
\setlength{\tabcolsep}{10pt}
\renewcommand{\arraystretch}{1.3}
\begin{tabular}{|c|c|c|c|c|}
\hline
\textbf{Matter type} & \(\omega\) & \(\kappa\) sign & Temperature \(T = -\kappa / 2\pi\) & \textbf{Horizon type} \\
\hline
Inflation & \(-1\) & \(\kappa < 0\) & \(T > 0\) & Past–inner (radiative) \\
\hline
Dust (Matter) & \(0\) & \(\kappa < 0\) & \(T > 0\) & Past–inner (radiative) \\
\hline
Radiation & \(1/3\) & \(\kappa < 0\) & \(T > 0\) & Past–inner (radiative) \\
\hline
Stiff matter & \(1\) & \(\kappa < 0\) & \(T > 0\) & Past–inner (radiative) \\
\hline
\end{tabular}
\caption{Behavior of the surface gravity \(\kappa\), temperature, and horizon type for different cosmic fluids in the early universe (\(H^2 \gg 1\)) with exponential entropy correction parameter \(\alpha = -1\).}
\label{table4}
\end{table}

At this stage, it is pertinent to make some comments in light of the above discussion and the summaries in tables \ref{table3} and \ref{table4}. Clearly, the case of negative values of the parameter $\alpha<0$ (table \ref{table4}) is more feasible than the information presented in table \ref{table3}, in light of previous studies \cite{Cai:2006rs}. This is because an outer $AH$ requires the presence of collapsing regions, or gravitational wells, where ingoing null rays diverge. In a perfectly homogeneous and isotropic FLRW universe, both at early and late times, the expansion of spacetime prevents the formation of such regions. Consequently, the only possible marginally trapped surface in an FLRW universe is an inner $AH$, located at the Hubble's radius. Early-time strong expansion and late-time accelerated expansion both reinforce this conclusion, making outer horizons physically implausible. However, when the dynamics of the Universe's evolution are altered by introducing new terms that change the form of the Friedmann equations, the possibility of a richer scenario (at least theoretically) increases, with the underlying reasons being supported by geometric, dynamical and thermodynamic arguments.

Now the picture is clear about all possibilities in having positive or negative temperatures, putting together (\ref{eq13}) and (\ref{eq14}), one obtains the following EoS,
\begin{equation}
P=\frac{3}{8 \pi R_{AH}^2}\left\{1-\alpha\left[e^{-\pi R_{AH}^2}+\pi R_{AH}^2 E i\left(-\pi R_{AH}^2\right)\right]\right\}   +\frac{1}{8 \pi} \dot{H}\left\{1-\alpha e^{-\pi R_{AH}^2}\right\}.
\end{equation}
Next, using (\ref{SG}) and (\ref{eq12}) to eliminate $\dot{H}$ in terms of $T$, one gets 
\begin{equation}\label{EoSvt}
P(v,T)=\frac{\left(1-2\pi Tv\right)\left(1-\alpha e^{-\pi v^{2}/4}\right)}{2\pi v^{2}}-\frac{3}{8}\alpha E i\left(-\pi v^{2}/4\right),
\end{equation}
being $v=2R_{AH}$ the reduced volume. It is clear that $\alpha\rightarrow 0$ leads to the pure GR EoS in a FLRW background. As it is well known, this scenario does not exhibit critical phenomena. Therefore, in the present context, it is expected that non-trivial thermodynamic phenomena will occur, with the exponential corrections to the entropy playing a significant role. 
To begin the study of the thermal stability of this model, we compute the heat capacity from Eq. (\ref{EoSvt}) and move toward understanding the possible critical behavior of this modified EoS for an entropic cosmological model with an exponential correction to the entropy. The behavior of the heat capacity $C_P$ provides important insights into the thermodynamic properties of the apparent horizon. 
The change of sign of $C_P$ distinguishes stable regimes ($C_P > 0$), where the system reacts smoothly to energy exchange, 
from unstable ones ($C_P < 0$), where thermodynamic equilibrium cannot be maintained. 
This sign flip is directly controlled by the deformation parameter $\alpha$, which therefore plays the role of a stability selector. 

In addition, the divergence of $C_P$ indicates the presence of a critical point in the system, 
marking the onset of a phase transition in the horizon thermodynamics. 
At this point, small perturbations in the horizon radius can give us large variations in energy transfer, 
reflecting a change in the thermodynamic phase of the Universe. Thus, the combined effect of the sign change and divergence of the heat capacity highlights the
existence of distinct thermodynamic phases of the horizon, 
with stability or instability governed by the interaction between geometric corrections (encoded in $\alpha$) 
and the dynamics of cosmic expansion. In this regard, Fig. \ref{figCp} shows an interesting situation for the chosen values of the parameter $\alpha$. On one hand, for $\alpha>0$ the system seems to have a phase transition stable in nature (red line), while for $\alpha<0$ there is a more involved situation: i) a stable phase transition (continuous blue line) and ii) an unstable phase transition, nonphysical (dashed blue line). 

Although a full thermodynamic stability scheme contemplates the study of both $C_{P}$ and $C_{v}$. In this case, $C_{v}=0$, then it is not possible to extract any information from this quantity. The fact of having a null $c_{v}$ for the present case is based on the type of phase transition suffered by the system. For the present case, we have a first-order phase transition, similar to the Van der Waals liquid-gas transition. Thus, the entropy is not a function of the temperature, leading to $C_{v}=0$. Therefore, at first sight, the only indicator for stable/unstable thermodynamic states lies in $C_{P}$.\footnote{Considering this point, it is worth mentioning that, in references \cite{Saha:2024mwn,
Bhandari:2017cow} has been explored the thermodynamics stability using a pure geometrical approach based on the generalized second law, deceleration parameter $q$ and both $C_{P}$ and $C_{v}$. However, this approach does work for the present situation because the thermodynamic situation is more involved leading to $C_{v}=0$. } These facts will be clear in the next section.

\begin{figure}[H]
    \centering
\includegraphics[width=0.68\textwidth]{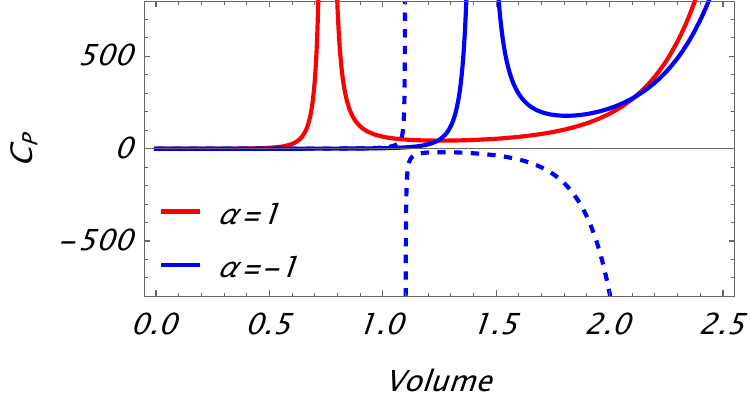}\ 
    \caption{Heat capacity $C_P$ as a function of the reduced volume $v$ 
    for two representative values of the deformation parameter, $\alpha = 1$ and $\alpha = -1$. 
    The sign of $\alpha$ controls the thermodynamic behavior of the system: 
    positive $\alpha$ leads to a stable branch with positive heat capacity, 
    while a negative $\alpha$ from one side introduces regions of negative heat capacity, 
    signaling possible thermodynamic instabilities. Besides, at points where the heat capacity diverges, a phase transition occurs.  }
    \label{figCp}
\end{figure}

\section{Phase transitions}\label{sec4}

To fully understand thermodynamic critical phenomena, it is important to determine whether the system undergoes phase transitions. This implies studying critical points that satisfy the following relations.
\begin{equation}\label{criticality}
    \left(\frac{\partial P}{\partial v}\right)\bigg{|}_{T}= \left(\frac{\partial^{2} P}{\partial v^{2}}\right)\bigg{|}_{T}=0.
\end{equation}
As (\ref{EoSvt}) is an intricate expression, the criticality conditions (\ref{criticality}) will be too. Consequently, it is not possible to obtain analytical solutions for $\{v,T\}$ from the system (\ref{criticality}). To proceed, we solve this system numerically for the tracking parameter $\alpha$ with values 1 and -1. These values respect the conditions discussed in the previous section. For $\alpha=1$, the system has only one critical point with $T<0$. The left panel of Fig. \ref{fig1} shows the trend of pressure as a function of reduced volume. As can be observed for small volumes, the pressure increases; however, for sufficiently large volumes, it decreases significantly. However, there is a region where $\partial p/\partial v > 0$, indicating that the system undergoes a first-order phase transition. Interestingly, this phenomenon is observed for temperatures $T<T_{c}$ above the critical isotherm, contrary to what happens in usual first-order phase transitions, where it appears below the critical isotherm. On the other hand, for the case $\alpha<-1$ the situation is more involved. Here, there is a double criticality. For the former, the only critical point leads to the usual small-to-large first-order phase transition as depicted in Fig. \ref{fig2} where the common swallowtail appears. In this case, the phenomenon is physical, as the Gibbs potential reaches its minimum. On the other hand, for $\alpha<0$, the Gibbs potential exhibits a more complex behavior. In Fig. \ref{fig3}, we show a number of snapshots for a relevant range of $T$ for $\alpha =- 1$. At $T_{c1}$, the first critical point appears, after which a physical swallowtail emerges, providing a local minimum for the Gibbs potential. However, as the temperature increases further in magnitude, this physical swallowtail moves leftward and, eventually, at $T\approx 0.07981$, it intersects the upper part of the curve, that is, the unstable branch (see the middle panel in the top row of Fig. \ref{fig3}). For $T_{f}$, we observe a zero-order phase transition (see dashed black line) that consists of a jump in the value of the thermodynamic potential. This phase transition starts from $T\approx 0.07981$ and extends until $T=T_{f}$, when the pressure of the left parts of the swallowtail coincides (see left panel in the middle row of Fig. \ref{fig3}). Beyond this it moves upward as temperature further increases until intersects the lower part of the physical swallowtail, shown in the right panel in the middle row of Fig. \ref{fig3}, after which it disappears, emerging an nonphysical swallowtail which continues to shrink as temperature increases until the second critical point $T_{c2}\approx 0.11037$ where it disappears. Then there is no critical behavior, leaving a cuspy Gibbs potential that displays unstable behavior. All these situations contribute to the existence of a reentrant phase transition between stable and unstable states.   

\begin{figure}[H]
    \centering
\includegraphics[width=0.48\textwidth]{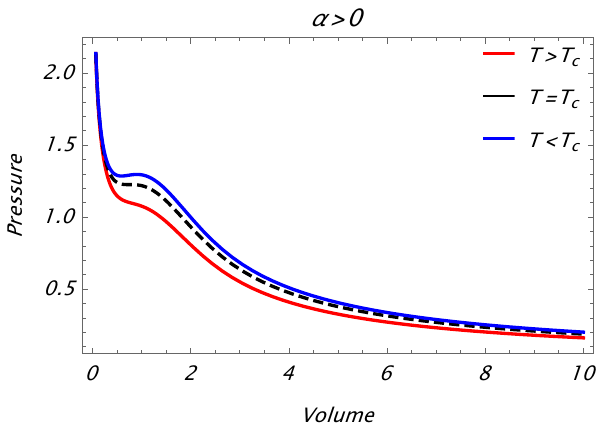}\ \includegraphics[width=0.48\textwidth]{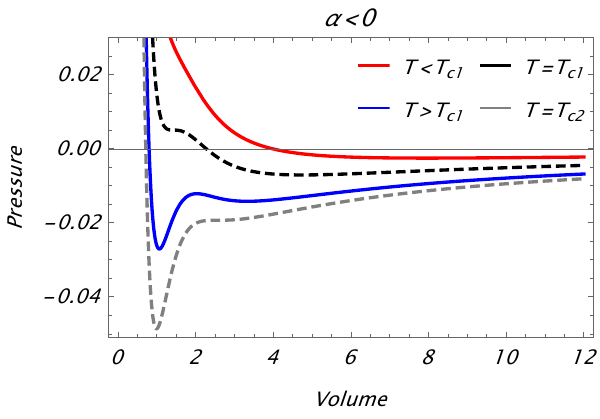} 
    \caption{The trend of the pressure versus the reduced volume for different values of the temperature, considering $\alpha=1$ (left panel) and $\alpha=-1$ (right panel). }
    \label{fig1}
\end{figure}

\begin{figure}[H]
    \centering
\includegraphics[width=0.48\textwidth]{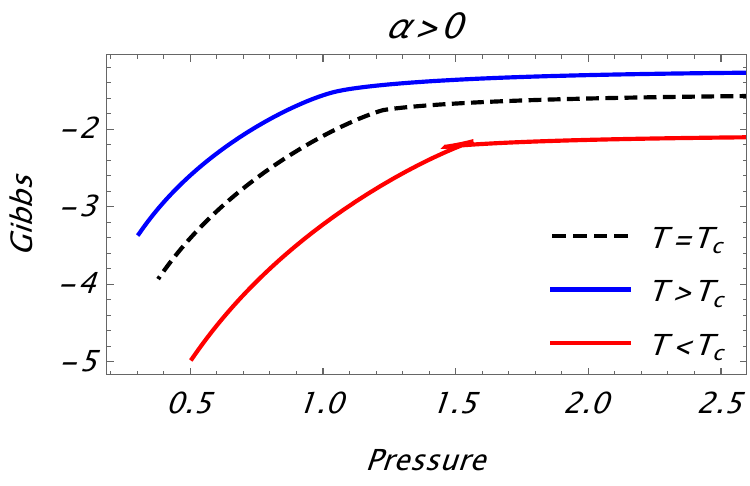}\ \includegraphics[width=0.48\textwidth]{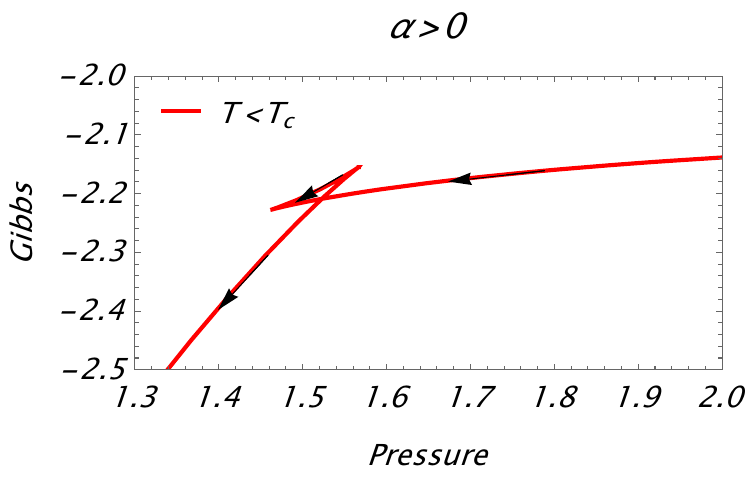} 
    \caption{ The behavior of the Gibbs free energy versus the pressure for increasing values of the temperature $T$.    }
    \label{fig2}
\end{figure}

\begin{figure}[H]
    \centering
\includegraphics[width=0.32\textwidth]{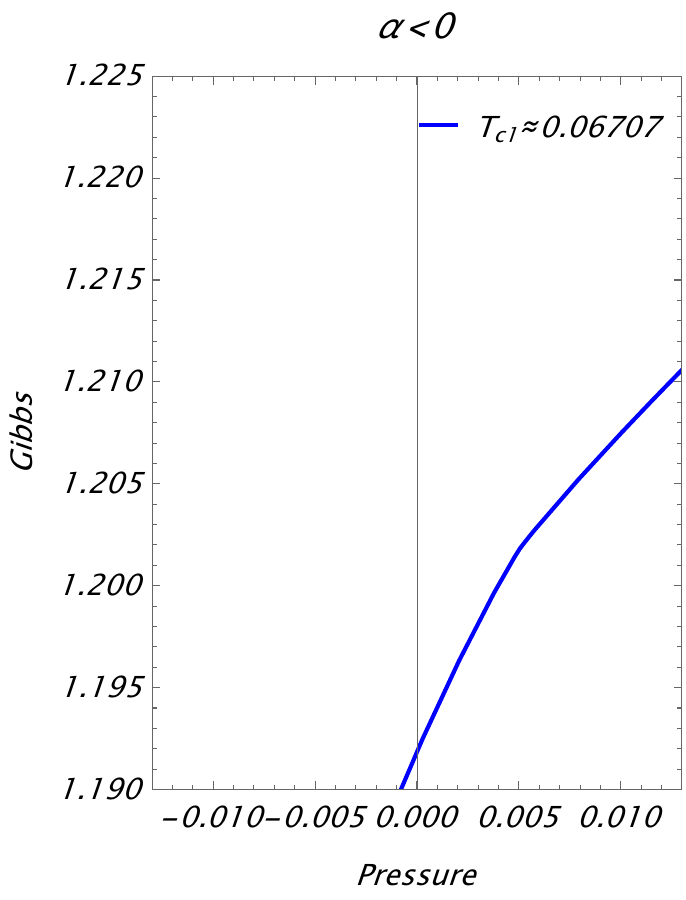}\ \includegraphics[width=0.32\textwidth]{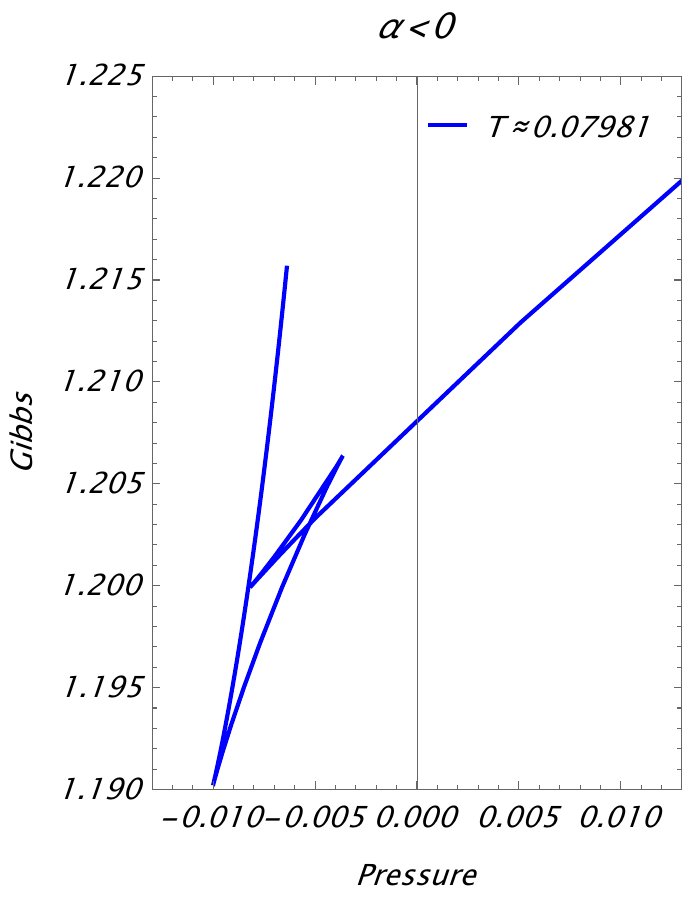} \ \includegraphics[width=0.32\textwidth]{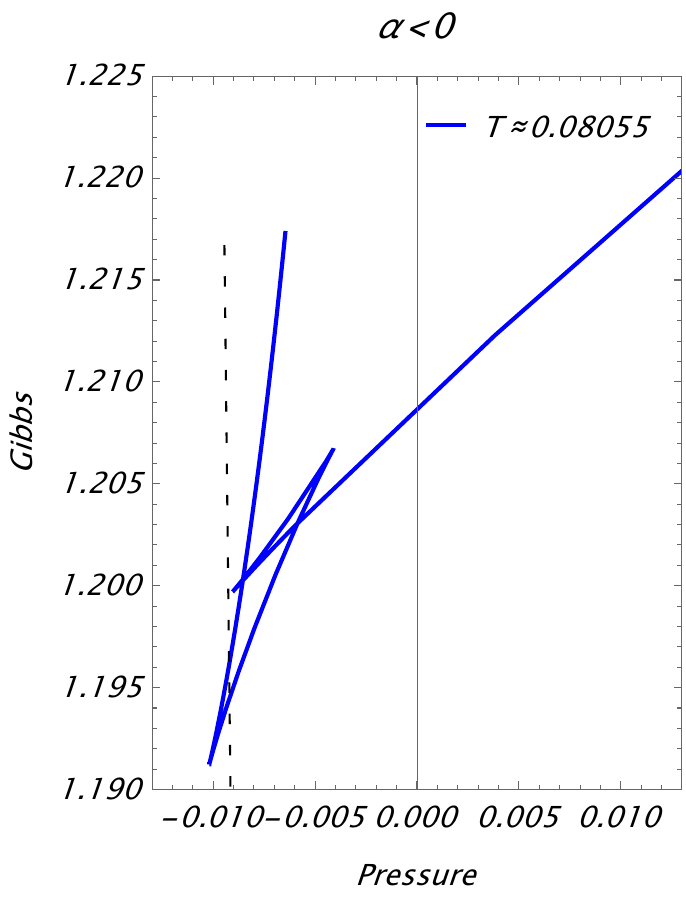} \\
\includegraphics[width=0.32\textwidth]{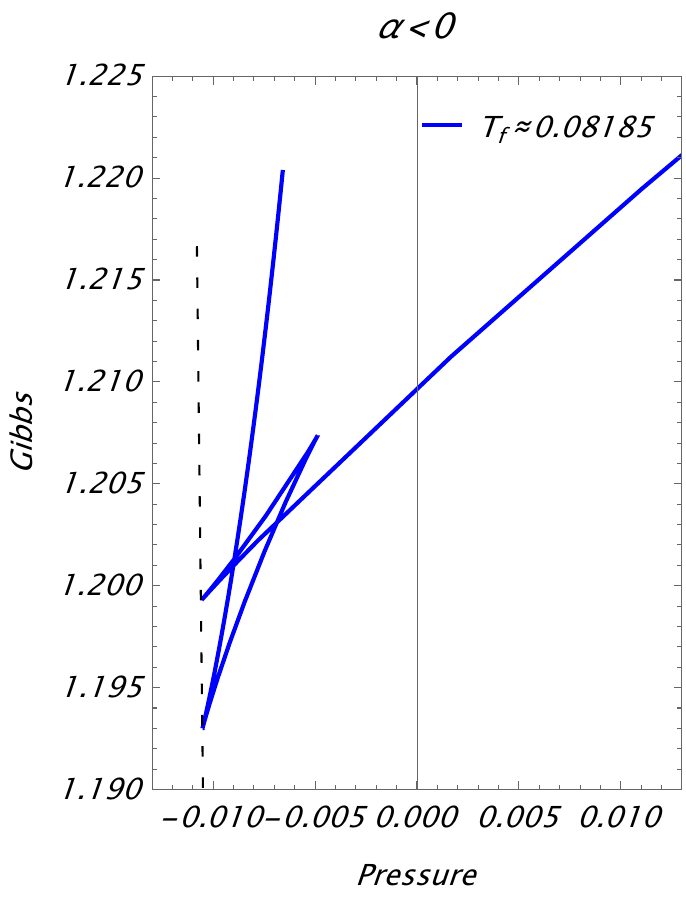}\ \includegraphics[width=0.32\textwidth]{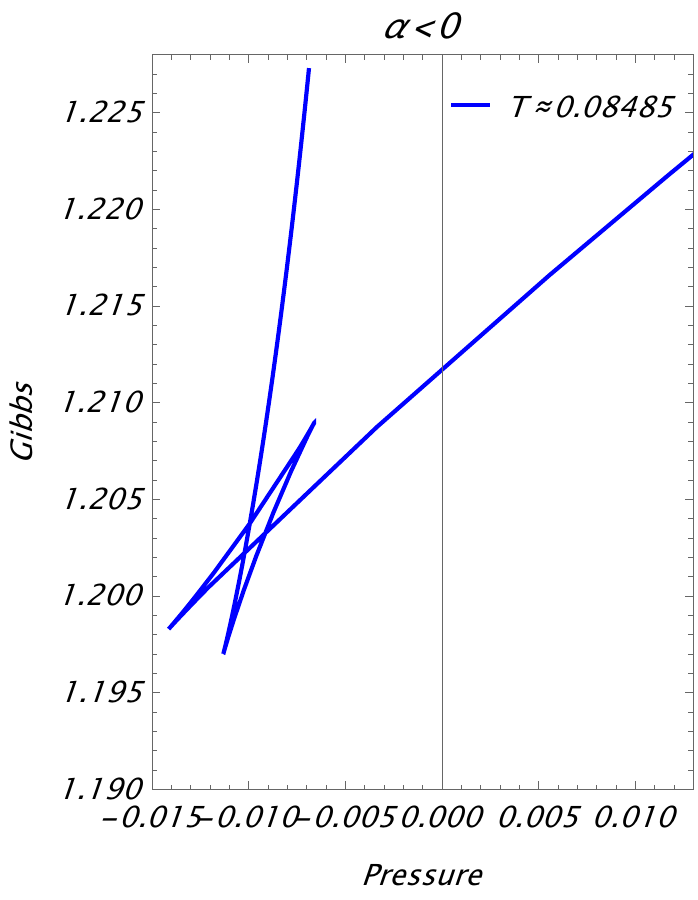}\ \includegraphics[width=0.32\textwidth]{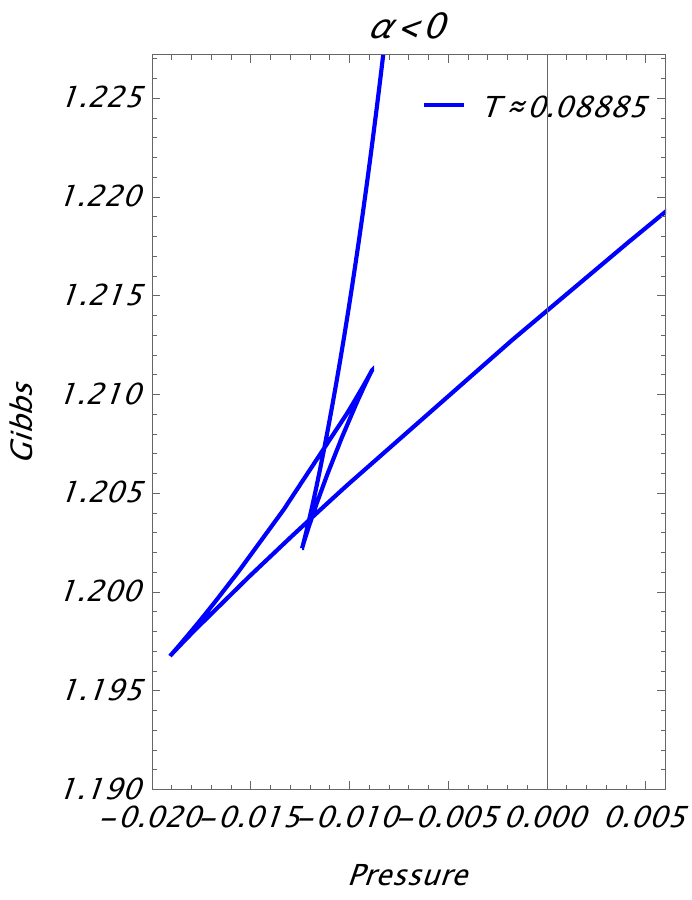}\\ 
\includegraphics[width=0.32\textwidth]{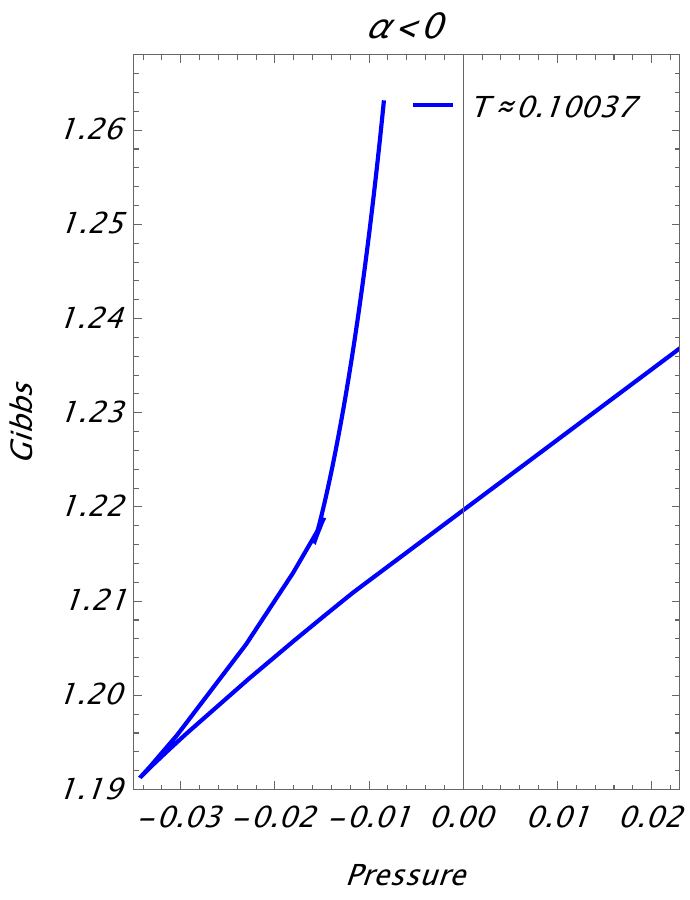}\ 
\includegraphics[width=0.32\textwidth]{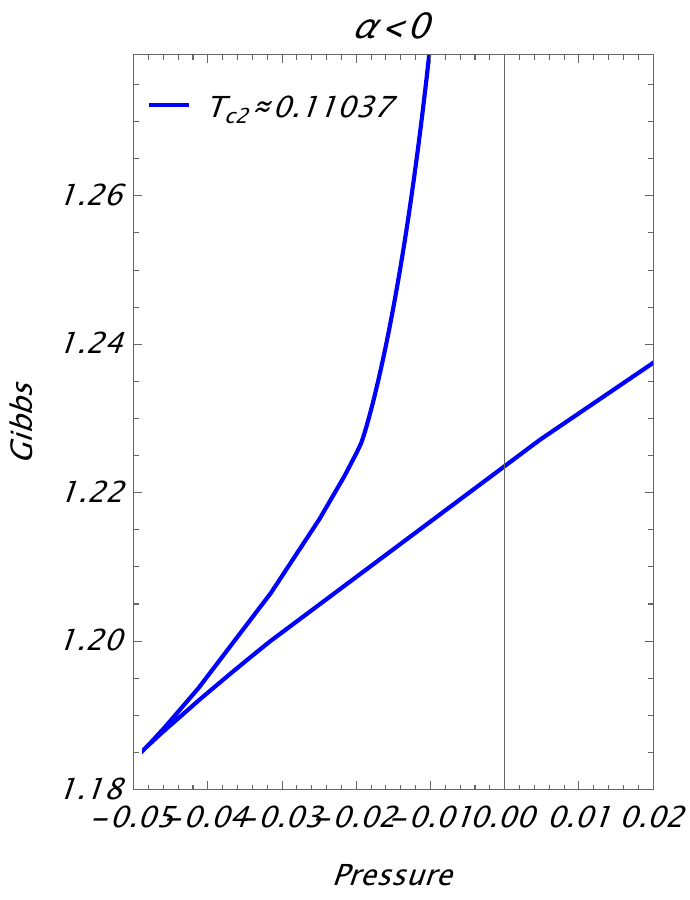}\ 
\includegraphics[width=0.32\textwidth]{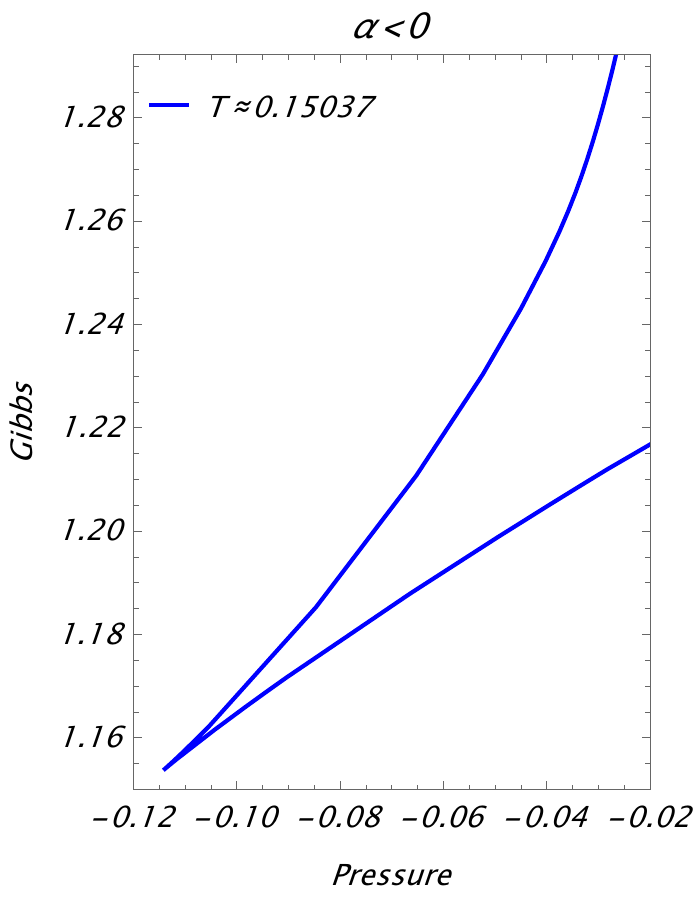}
    \caption{ The behavior of the Gibbs free energy versus the pressure for increasing values of the temperature $T$.   }
    \label{fig3}
\end{figure}

\section{A glance of entropic cosmology for exponential correction of the entropy}\label{sec5}

In this section, we aim to explore the cosmological implications of our entropic model and its comparison with observational data. First, we will start with the continuity equation, which represents energy conservation in our context. 
\begin{equation}
\dot{\rho} + 3H(\rho + p) = 0, \label{eqrho}
\end{equation}
where we consider a barotropic fluid described by $p = \omega \,\rho$ as the equation of state. From eq. (\ref{eqrho}) We can obtain the evolution of density in terms of the cosmological redshift as follows
\begin{align}
\rho(z) &= \rho_0 \,(1+z)^{3(1+w)}.
\label{eq:rho_of_z}
\end{align}
In the next step, we are adding the cosmological constant term to the modified Friedmann equation as follows,
\begin{equation}
    H^2\Bigg\{1
- \alpha\!\left[
e^{-\frac{\pi}{H^2}}
+ \frac{\pi}{H^2}\,\mathrm{Ei}\!\!\left(-\frac{\pi}{H^2}\right)
\right]\Bigg\}= \frac{8\pi G}{3}\,\rho(z) + \frac{\Lambda}{3}. \label{modFRI1}
\end{equation}
In order to do the comparison with the $\Lambda$CDM models we are considering a pressureless matter ($\Omega =0$) and the density parameters for dark matter $\Omega_m$ and dark energy $\Omega_\Lambda$
\begin{align}
\rho(z) &= \rho_{m0}\,(1+z)^3,
\qquad
\rho_{m0} \;=\; \frac{3H_0^2\,\Omega_m}{8\pi G},
\qquad
\frac{\Lambda}{3} \;=\; H_0^2\,\Omega_\Lambda,
\label{eq:rhom_lambda_defs}
\end{align}
such that the Friedmann equation in terms of the cosmological redshifts and the cosmological parameters is given by
\begin{align}
H^2(z)\Bigg\{1
- \alpha\!\left[
e^{-\frac{\pi}{H^2(z)}}
+ \frac{\pi}{H^2(z)}\,\mathrm{Ei}\!\!\left(-\frac{\pi}{H^2(z)}\right)
\right]\Bigg\}
&= H_0^2\!\left[\Omega_m(1+z)^3 + \Omega_\Lambda\right],
\label{H(z)}
\end{align}
here we see that for milit case $\alpha=0$ we recover the $H(z)$ for the $\Lambda$CDM model
\begin{align}
H^2(z) \;=\; H_0^2\!\left[\Omega_m(1+z)^3 + \Omega_\Lambda\right].
\label{Lcdm}
\end{align}

Eq. (\ref{H(z)}) is an implicit equation that defines $H(z)$ and for each $z$ value we need to numerically solve for $H(z)$ imposing the condition that $H(0)=H_0$. The main results regarding this point are summarized in Fig. \ref{fig5}, where we can see that the positive $\alpha$ case leads to faster growth of $H(z)$ at high redshifts, while the negative $\alpha$ case predicts lower expansion rates. Also, we plot the distance modulus $\mu(z)$ from our model, and we compare with Pantheon+ SHOES compilation \cite{Scolnic:2021amr, Brout:2022vxf}; in both cases, we consider Planck 2018 parameters $H_0 = 67.4\ \text{km/s/Mpc}, \Omega_m = 0.315, \Omega_\Lambda = 0.685$.

The comparison between Pantheon+ SH0ES supernovae data and theoretical predictions reinforces the success of the $\Lambda$CDM model\footnote{In \cite{Mukherjee:2025xuy} a dynamical system approach has been used to explore the thermodynamic effects of $\Lambda$CDM model.}, calibrated with Planck 2018 parameters, in describing the late-time cosmic expansion. The $\Lambda$CDM curve (red dashed) closely follows the observed distance modulus across the redshift range $0 < z < 2$, lying well within the observational uncertainties.

However, the inclusion of a deformation parameter $\alpha$ through the exponential correction for the entropy in the modified cosmological model introduces systematic deviations from the standard expansion history. For $\alpha > 0$, the predicted distance modulus is shifted upward at intermediate redshifts, corresponding to a reduced expansion rate in the past. Conversely, negative values ($\alpha < 0$) yield lower distance moduli, consistent with a more rapid past expansion. These shifts are within the sensitivity of current supernova datasets, making Pantheon+ an effective probe of departures from $\Lambda$CDM (see Fig. \ref{fig6}).

Importantly, such modifications can affect the ongoing tension with $H_0$. The SH0ES calibration yields $H_0 \approx 73\ \text{km s}^{-1}\text{Mpc}^{-1}$, significantly higher than the Planck 2018 value of $H_0 \approx 67.4\ \text{km s}^{-1}\text{Mpc}^{-1}$. Positive values of $\alpha$, which reduce the effective expansion rate at higher redshifts, could potentially ease this discrepancy by reconciling the local and early-universe determinations of $H_0$. Conversely, negative $\alpha$ values would exacerbate the tension.

Thus, the analysis highlights that even minor departures from $\Lambda$CDM may leave detectable imprints in the supernova Hubble diagram, and that constraints on $\alpha$ from Pantheon+ and complementary probes (BAO, CMB, and cosmic chronometers) are crucial to assess whether such models can alleviate the $H_0$ tension or if the discrepancy points to more fundamental new physics.

\begin{figure}[H]
    \centering
\includegraphics[width=0.6\textwidth]{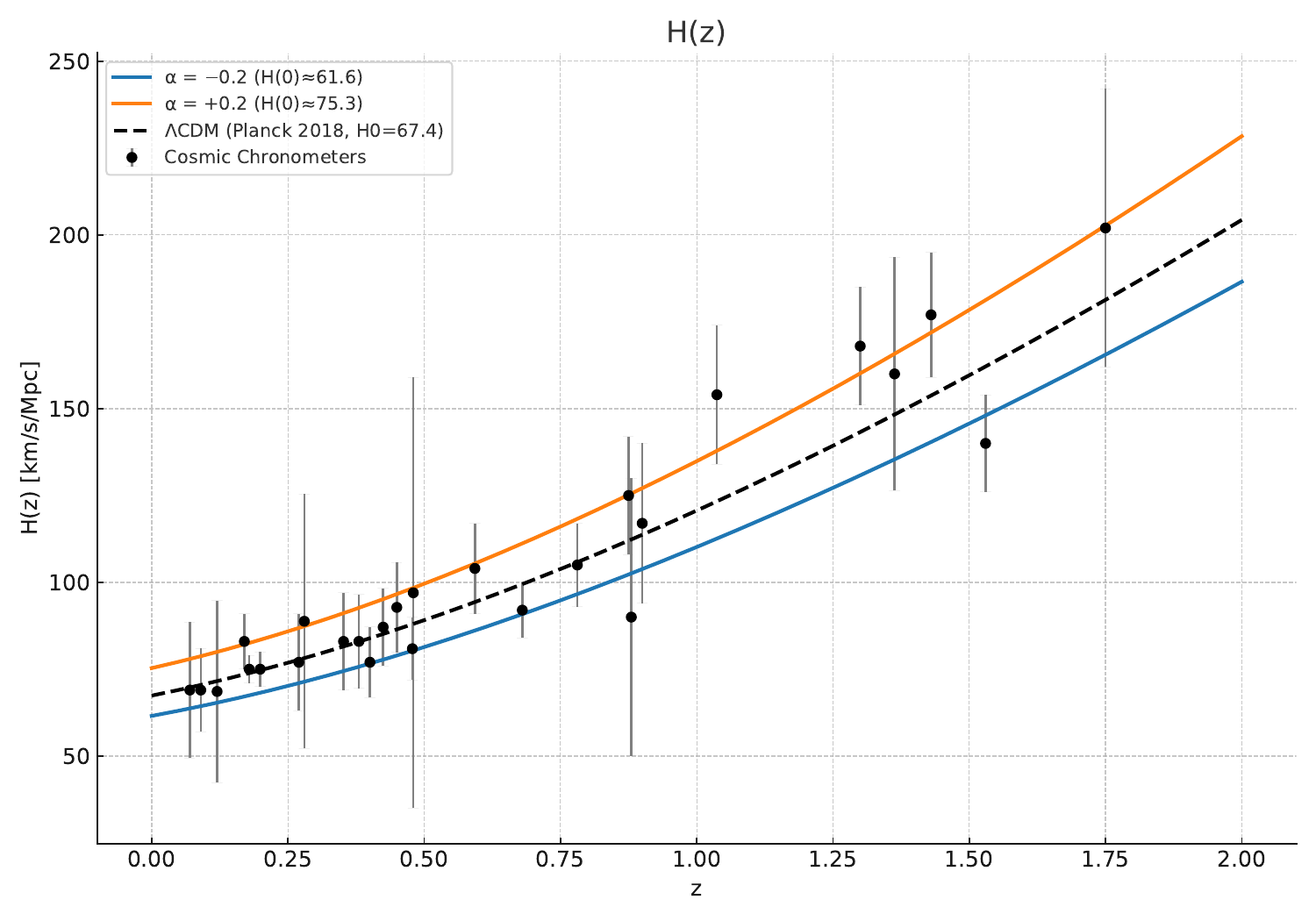}\ 
    \caption{ This plot show a comparison of the modified Friedmann equation (\ref{H(z)}) in our model with $\alpha=\pm 2$ (solid line) and the standard $\Lambda$CDM prediction from Planck 2018 parameters (black dashed line) against observational data from cosmic chronometers (black point) \cite{Jimenez2003, Simon2005, Stern2010, Moresco2012, Zhang2014, Moresco2015, Moresco2016, Ratsimbazafy2017}. The positive $\alpha$ case leads to a faster growth of $H(z)$ at high redshifts, while negative $\alpha$ case predicts lower expansion rates}
    \label{fig5}
\end{figure}

\begin{figure}[H]
    \centering
\includegraphics[width=0.6\textwidth]{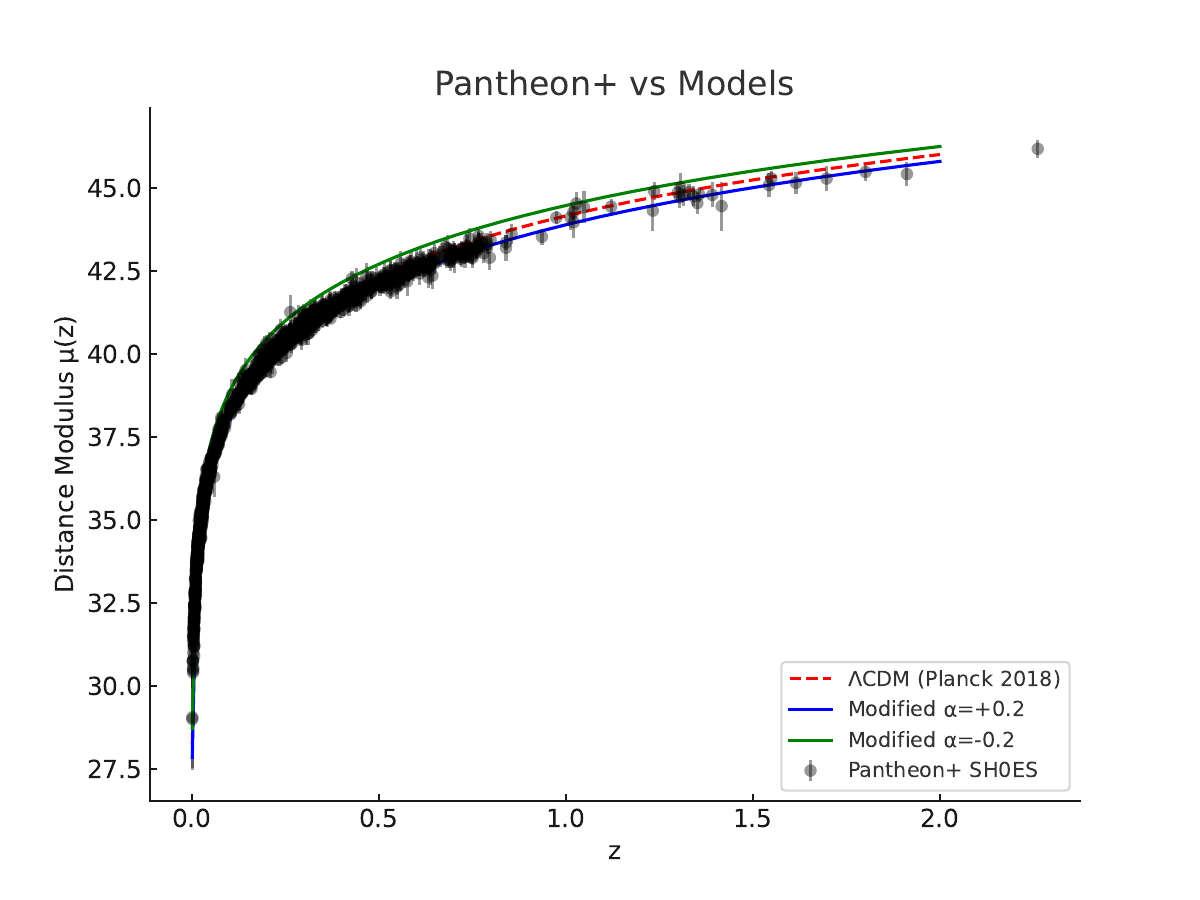}\ 
    \caption{
     Comparison of the distance modulus $\mu(z)$ from the Pantheon+ SH0ES compilation (black points) with theoretical predictions. The red dashed line corresponds to the standard $\Lambda$CDM model with Planck 2018 parameters ($H_0 = 67.4\ \text{km/s/Mpc}, \Omega_m = 0.315, \Omega_\Lambda = 0.685$). The solid blue and green curves represent the predictions of the modified cosmological model with deformation parameter values $\alpha = +0.2$ and $\alpha = -0.2$, respectively. The comparison illustrates the effect of $\alpha$ on the expansion history relative to $\Lambda$CDM.}
    \label{fig6}
\end{figure}

\subsection{Statistical methodology and model selection criteria}
\label{sec:AICBIC}
In this section, we summarize the statistical procedure used to confront the theoretical models with
the observational data and to evaluate their relative statistical performance.
The comparison between theoretical predictions and cosmological observations is quantified
through the $\chi^2$--square statistic,
\begin{equation}
\chi^2(\boldsymbol{\theta}) =
\sum_{i=1}^{N}
\frac{\big[y_{\mathrm{obs},i} - y_{\mathrm{th}}(z_i;\boldsymbol{\theta})\big]^2}
{\sigma_i^2} ,
\label{eq:chi2_def}
\end{equation}
where $y_{\mathrm{obs},i}$ denotes the observed quantity (either $H(z)$ or the distance modulus
$\mu(z)$), $\sigma_i$ is its associated uncertainty, and
$y_{\mathrm{th}}(z_i;\boldsymbol{\theta})$ is the theoretical prediction
obtained from the model parameters
$\boldsymbol{\theta} = (H_0, \Omega_m, \alpha)$.
The optimal parameters are found by minimizing $\chi^2(\boldsymbol{\theta})$,
and the minimum value $\chi^2_{\min}$ provides a direct measure of the overall goodness of fit.
A satisfactory cosmological model typically yields $\chi^2_{\min}/N_{\mathrm{dof}} \simeq 1$,
where $N_{\mathrm{dof}} = N - k$ is the number of degrees of freedom and $k$ is the number of free parameters.

Since the Pantheon+\,SH0ES supernova sample and the cosmic--chronometer
(CC) datasets are statistically independent,
their total likelihood can be expressed as a product
$\mathcal{L}_{\text{tot}} = \mathcal{L}_{\mu} \mathcal{L}_{H}$,
which is equivalent to an additive combination of their $\chi^2$--square contributions,
\begin{equation}
\chi^2_{\text{tot}} = \chi^2_{\mu} + \chi^2_{H} .
\label{eq:chi2_total}
\end{equation}
Here $\chi^2_{\mu}$ refers to the 1702 data points of the Pantheon+\,SH0ES compilation
and $\chi^2_H$ corresponds to the 31 CC measurements, yielding a total of
$N_{\text{tot}} = 1733$ independent observations.

Once the minimum chi--square values are obtained,
we evaluate the relative performance of competing models through the Akaike (AIC) and Bayesian
Information Criteria (BIC), defined respectively as
\begin{equation}
\mathrm{AIC} = \chi^2_{\min} + 2k , \qquad
\mathrm{BIC} = \chi^2_{\min} + k\ln N_{\text{tot}} ,
\label{eq:AICBIC_def}
\end{equation}
where $k$ is the number of free parameters.
These quantities penalize models with excessive complexity
and provide an objective basis for model comparison.

For the flat $\Lambda$CDM model we have $k_{\Lambda}=2$ ($H_0$, $\Omega_m$),
while the exponential--entropy model introduces one additional parameter
$k_{\exp}=3$ ($H_0$, $\Omega_m$, $\alpha$).
The relative differences,
\begin{equation}
\Delta\mathrm{AIC} = \mathrm{AIC}_{\exp}-\mathrm{AIC}_{\Lambda{\rm CDM}},
\qquad
\Delta\mathrm{BIC} = \mathrm{BIC}_{\exp}-\mathrm{BIC}_{\Lambda{\rm CDM}},
\label{eq:DeltaAICBIC}
\end{equation}
quantify the statistical preference between the two cosmologies.
In general, $\Delta\mathrm{AIC}<2$ indicates statistically equivalent models,
while $\Delta\mathrm{BIC}>2$ provides moderate evidence
in favor of the simpler model with the smaller BIC.

The joint analysis of the Pantheon+\,SH0ES supernova compilation and the
cosmic--chronometer (CC) dataset was performed using the $\chi^2$ minimization
procedure described below.
The best--fit cosmological parameters for each model were obtained by minimizing
Eq.~(\ref{eq:chi2_def}) with respect to the free parameters
$\boldsymbol{\theta} = (H_0, \Omega_m, \alpha)$.
We summarize the resulting values and the corresponding
minimum $\chi^2$--square statistic as follows

\begin{equation}
\begin{aligned}
\Lambda\mathrm{CDM:}\quad &
H_0 = 73.0^{+1.1}_{-1.0}\ \mathrm{km\,s^{-1}\,Mpc^{-1}}, &
\Omega_m = 0.35^{+0.03}_{-0.04}, &
\chi^{2\,\,\, (\Lambda\mathrm{CDM})}_{\min}=848.0, \nonumber\\[3pt]
\text{Exp. model:}\quad &
H_0 = 73.4^{+1.2}_{-1.0}\ \mathrm{km\,s^{-1}\,Mpc^{-1}}, &
\Omega_m = 0.34^{+0.03}_{-0.03}, &
\alpha = -0.06^{+0.08}_{-0.07}, \\&
\chi^{2(\exp)}_{\min}=847.5.
\label{eq:BestFitValues}
\end{aligned}
\end{equation}

Both models provide a good description of the combined data with
reduced $\chi^2$--square values close to unity
($\chi^2_{\min}/N_{\mathrm{dof}}\!\simeq\!1$),
demonstrating the internal consistency of the fit.
The exponential--entropy model reproduces the expansion history of
the concordance $\Lambda$CDM cosmology with almost identical accuracy.
The best--fit value $\alpha\simeq -0.06$ corresponds to a small negative
entropic correction that slightly enhances the late--time Hubble rate,
producing a marginally higher value of $H_0$.
This feature may help to alleviate the current $H_0$ tension
without compromising the fit to high--redshift observations.

Using the best--fit parameters derived previously,
we obtain $\chi^{2 \,\,\,(\Lambda{\rm CDM})}_{\min}\!\approx\!848$
and $\chi^{2\,\,\, (\exp)}_{\min}\!\approx\!847.5$ when both datasets are combined.
Consequently,
\begin{align}
\mathrm{AIC}_{\Lambda{\rm CDM}} &= 852.0, &
\mathrm{AIC}_{\exp} &= 853.5, &
\Delta\mathrm{AIC} &\simeq 1.5 , \nonumber\\
\mathrm{BIC}_{\Lambda{\rm CDM}} &\simeq 862.9, &
\mathrm{BIC}_{\exp} &\simeq 869.9, &
\Delta\mathrm{BIC} &\simeq 7.0 .
\label{eq:AICBIC_values}
\end{align}
These values show that the exponential--entropy cosmology reproduces
the expansion history of $\Lambda$CDM with excellent accuracy
and remains observationally viable,
although it is not statistically favored once the additional parameter is penalized.
Nevertheless, small negative values of $\alpha$
slightly increase $H_0$ at low redshift,
suggesting that the entropic correction can soften the present $H_0$ tension
without compromising consistency with the Pantheon+\,SH0ES and CC data.
These values show that the exponential--entropy cosmology reproduces
the expansion history of $\Lambda$CDM with high accuracy
and remains observationally viable,
although it is not statistically preferred once the additional parameter is penalized.
Nevertheless, small negative values of $\alpha$ slightly raise $H_0$ at low redshift,
suggesting that the entropic correction can mildly alleviate the current $H_0$ tension
without spoiling the overall consistency with the Pantheon+\,SH0ES and
cosmic--chronometer data

The statistical analysis carried out in this section shows that the exponential--entropy
cosmology provides an excellent fit to the combined Pantheon+\,SH0ES, and cosmic--chronometer
datasets, with a minimum chi--square value essentially indistinguishable from that of the
standard $\Lambda$CDM scenario.
When evaluated through model--selection criteria, the Akaike Information Criterion indicates
that both models offer statistically equivalent descriptions of the data, whereas the Bayesian
Information Criterion moderately favors $\Lambda$CDM due to its smaller number of free parameters.
Overall, the results demonstrate that introducing the exponential entropy correction does not
degrade the agreement with observations and remains fully compatible with the current expansion
history of the Universe.
This establishes the exponential--entropy framework as a viable and observationally consistent
extension of the standard cosmological model, with potential physical implications to explore in future work

\section{Concluding remarks}\label{sec6}
In this work we have explored how exponential corrections to the horizon entropy modify the thermodynamic and geometric behavior of the apparent horizon in a flat FLRW universe. By employing the unified first law of thermodynamics together with the Hayward--Kodama surface gravity, we derived the modified Friedmann equations and constructed the corresponding effective equation of state. These corrections generate a rich thermodynamic structure that is absent in standard GR cosmology, including divergences in the heat capacity, critical points, and first-order phase transitions.

A central aspect of our analysis is the dependence of the temperature’s sign on the causal character of the apparent horizon. Since FLRW is homogeneous and isotropic at both early and late times, its apparent horizon is necessarily a \textit{past-inner} marginal surface, regardless of the matter content considered. This geometric restriction fixes the sign of $\kappa_{\mathrm{HK}}$ and, therefore, the sign of the temperature. As a consequence, although the modified Friedmann equations formally allow regimes with $\kappa_{\mathrm{HK}}>0$ (and thus $T<0$), these correspond to \textit{past-outer} horizons, which are incompatible with the causal structure of FLRW spacetime. Only the inner-horizon configurations (and therefore the corresponding thermodynamic branches) represent physically meaningful scenarios. This observation naturally excludes some of the mathematically admissible branches in the $P$-$v$ diagram.

From a thermodynamic viewpoint, exponential corrections introduce new stable and unstable regions, controlled by the deformation parameter $\alpha$. For $\alpha>0$, the system exhibits a single critical point and a first-order phase transition at negative temperature, occurring above the critical isotherm---a reversed scenario compared to standard black-hole thermodynamics. For $\alpha<0$, the system displays double criticality and a more intricate structure, including a reentrant phase transition and a zero-order jump in the Gibbs free energy. However, this second branch contains extended intervals in which $C_{P}<0$ and no physically acceptable minimum of the Gibbs potential exists, rendering it thermodynamically unstable. When the geometric constraint on the sign of $\kappa_{\mathrm{HK}}$ is combined with the thermodynamic stability conditions, only the $\alpha>0$ branch survives as a physically viable description of early-Universe horizon thermodynamics. However, as mentioned, this case belong to an \emph{past-outer} horizon.

On the other hand, numerical integration of the modified cosmological equations demonstrates that positive values of the deformation parameter $\alpha$ suppress the expansion rate, while negative values enhance it. The deviations from $\Lambda$CDM become significant at low redshifts ($z \lesssim 2$), where observational data are most sensitive. Regarding observational viability, we confronted the modified Friedmann dynamics with cosmic-chronometer data and the Pantheon+SH0ES supernova compilation. The model reproduces the global expansion history with accuracy comparable to $\Lambda$CDM. Nonetheless, small deformations ($|\alpha| \lesssim 0.2$) are still allowed observationally and may represent subleading corrections to the expansion history. Moreover, the exponential correction mildly shifts the inferred value of $H_{0}$ toward higher values, it does not lead to a statistically significant improvement once model-selection criteria (AIC/BIC) are applied, as $\Lambda$CDM remains favored due to the penalty associated with the additional parameter $\alpha$. Therefore, the fact that this entropy deformation yields a consistent background cosmology while simultaneously predicting non-trivial horizon thermodynamics is noteworthy.

Taken together, these statistical results place the exponential--entropy scenario on firm
observational grounds and justify examining its physical implications beyond the background
expansion. 
In particular, the presence of the non-perturbative exponential term $e^{-A/4}$ in the horizon
entropy induces qualitative modifications in the thermodynamic structure of the apparent horizon,
including the emergence of a $P$-$v$-type phase transition in the early Universe and a
reentrant phase transition for negative values of the deformation parameter $\alpha$. 
Such features have no analogue in the standard FLRW model with a perfect fluid, where the
The thermodynamic evolution is strictly monotonic, with no phase transitions.
Therefore, the observational viability established in this section provides the necessary
foundation for interpreting these thermodynamic effects as genuine signatures of the underlying
quantum-gravitational corrections encoded in the exponential entropy.

Overall, our results suggest that exponential corrections to the area law, while preserving the leading GR behavior at large scales, introduce rich thermodynamic features in the early Universe and select physically meaningful branches through the interplay between stability and horizon geometry. These findings highlight the importance of horizon thermodynamics as a probe of quantum-gravitational microstructure. The identification of phase transitions, critical points, and Gibbs energy instabilities suggests that the thermodynamics of the cosmic horizon is richer than in the standard picture. These features may leave subtle imprints on cosmological observables, including a possible alleviation of the current $H_0$ tension through entropy-induced corrections.

Future work may extend this analysis to anisotropic or inhomogeneous cosmologies, where the causal character of marginal surfaces is less constrained, allowing for a broader range of thermodynamic behavior and potentially new signatures of entropy deformations.

\section*{ACKNOWLEDGEMENTS}
We thank the anonymous referees for their enlightening comments. We thank Manuel Gonzalez-Espinoza for explaining the AIC and BIC criteria. J. Saavedra acknowledges the grant FONDECYT N°1220065, Chile.

\section*{Data Availability Statement}
There are no new data associated with this article.


\begin{thebibliography}{99}
\bibitem{Bekenstein:1973ur}
J.~D.~Bekenstein,
Phys. Rev. D \textbf{7}, 2333-2346 (1973)
doi:10.1103/PhysRevD.7.2333

\bibitem{Bardeen:1973gs}
J.~M.~Bardeen, B.~Carter and S.~W.~Hawking,
Commun. Math. Phys. \textbf{31}, 161-170 (1973)
doi:10.1007/BF01645742

\bibitem{Hawking:1975vcx}
S.~W.~Hawking,
Commun. Math. Phys. \textbf{43}, 199-220 (1975)
[erratum: Commun. Math. Phys. \textbf{46}, 206 (1976)]
doi:10.1007/BF02345020

\bibitem{Gibbons:1976ue}
G.~W.~Gibbons and S.~W.~Hawking,
Phys. Rev. D \textbf{15}, 2752-2756 (1977)
doi:10.1103/PhysRevD.15.2752

\bibitem{Jacobson:1995ab}
T.~Jacobson,
Phys. Rev. Lett. \textbf{75}, 1260-1263 (1995)
doi:10.1103/PhysRevLett.75.1260
[arXiv:gr-qc/9504004 [gr-qc]].

\bibitem{Kodama:1979vn}
H.~Kodama,
Prog. Theor. Phys. \textbf{63}, 1217 (1980)
doi:10.1143/PTP.63.1217

\bibitem{Hayward:1993wb}
S.~A.~Hayward,
Phys. Rev. D \textbf{49}, 6467-6474 (1994)
doi:10.1103/PhysRevD.49.6467

\bibitem{Hayward:1994bu}
S.~A.~Hayward,
Phys. Rev. D \textbf{53}, 1938-1949 (1996)
doi:10.1103/PhysRevD.53.1938
[arXiv:gr-qc/9408002 [gr-qc]].

\bibitem{Hayward:1997jp}
S.~A.~Hayward,
Class. Quant. Grav. \textbf{15}, 3147-3162 (1998)
doi:10.1088/0264-9381/15/10/017
[arXiv:gr-qc/9710089 [gr-qc]].

\bibitem{Cai:2005ra}
R.~G.~Cai and S.~P.~Kim,
JHEP \textbf{02}, 050 (2005)
doi:10.1088/1126-6708/2005/02/050
[arXiv:hep-th/0501055 [hep-th]].

\bibitem{Cai:2006pa}
R.~G.~Cai and L.~M.~Cao,
Nucl. Phys. B \textbf{785}, 135-148 (2007)
doi:10.1016/j.nuclphysb.2007.06.016
[arXiv:hep-th/0612144 [hep-th]].

\bibitem{Cai:2006rs}
R.~G.~Cai and L.~M.~Cao,
Phys. Rev. D \textbf{75}, 064008 (2007)
doi:10.1103/PhysRevD.75.064008
[arXiv:gr-qc/0611071 [gr-qc]].

\bibitem{Saha:2024mwn}
S. Saha, S. Saha and N. Mahata, Class. Quant. Grav. \textbf{42}, 055018 (2025) doi: 10.1088/1361-6382/adb2d3.

\bibitem{Bhandari:2017cow} P. Bhandari, S. Haldar, S. Chakraborty, Eur. Phys. J. C \textbf{77}, 840 (2017) doi:10.1140/epjc/s10052-017-5417-1.  

\bibitem{Li:2013fop}
H.~Li and Y.~Zhang,
Commun. Theor. Phys. \textbf{60}, 28-36 (2013)
doi:10.1088/0253-6102/60/1/05

\bibitem{Sebastiani:2023brr}
L.~Sebastiani,
Phys. Dark Univ. \textbf{42}, 101296 (2023)
doi:10.1016/j.dark.2023.101296
[arXiv:2307.04509 [gr-qc]].

\bibitem{Nojiri:2025gkq}
S.~Nojiri, S.~D.~Odintsov, T.~Paul and S.~SenGupta,
[arXiv:2503.19056 [gr-qc]].

\bibitem{Nojiri:2024zdu}
S.~Nojiri, S.~D.~Odintsov and T.~Paul,
Universe \textbf{10}, no.9, 352 (2024)
doi:10.3390/universe10090352
[arXiv:2409.01090 [gr-qc]].

\bibitem{Nojiri:2023wzz}
S.~Nojiri, S.~D.~Odintsov, T.~Paul and S.~SenGupta,
Phys. Rev. D \textbf{109}, no.4, 043532 (2024)
doi:10.1103/PhysRevD.109.043532
[arXiv:2307.05011 [gr-qc]].

\bibitem{Nojiri:2022nmu}
S.~Nojiri, S.~D.~Odintsov and T.~Paul,
Phys. Lett. B \textbf{835}, 137553 (2022)
doi:10.1016/j.physletb.2022.137553
[arXiv:2211.02822 [gr-qc]].


\bibitem{Bamba:2018zil} K. Bamba, A. Jawal, S. Rafique and H. Moradpour, Eur. Phys. J. C \textbf{78}, 986 (2018) 
doi:10.1140/epjc/s10052-018-6446-0.


\bibitem{Kaniadakis:2002zz}
G.~Kaniadakis,
Phys. Rev. E \textbf{66}, 056125 (2002)
doi:10.1103/PhysRevE.66.056125
[arXiv:cond-mat/0210467 [cond-mat.stat-mech]].

\bibitem{Kaniadakis:2005zk}
G.~Kaniadakis,
Phys. Rev. E \textbf{72}, 036108 (2005)
doi:10.1103/PhysRevE.72.036108
[arXiv:cond-mat/0507311 [cond-mat]].




\bibitem{Zhang:2008gt}
J.~Zhang,
Phys. Lett. B \textbf{668}, 353-356 (2008)
doi:10.1016/j.physletb.2008.09.005
[arXiv:0806.2441 [hep-th]].



\bibitem{Barrow:2020tzx}
J.~D.~Barrow,
Phys. Lett. B \textbf{808}, 135643 (2020)
doi:10.1016/j.physletb.2020.135643
[arXiv:2004.09444 [gr-qc]].
\bibitem{Jawad:2025uqu} A. Jawad, N. Azhar, Warisha, N. Myrzakulov, K. Yerzhanov and S. Myrzakul. High Energy Dens. Phys. \textbf{55}, 101187 (2025) 
doi: 10.1016/j.hedp.2025.101187



\bibitem{Nojiri:2022aof}
S.~Nojiri, S.~D.~Odintsov and V.~Faraoni,
Phys. Rev. D \textbf{105}, no.4, 044042 (2022)
doi:10.1103/PhysRevD.105.044042
[arXiv:2201.02424 [gr-qc]].

\bibitem{Nojiri:2022dkr}
S.~Nojiri, S.~D.~Odintsov and T.~Paul,
Phys. Lett. B \textbf{831}, 137189 (2022)
doi:10.1016/j.physletb.2022.137189
[arXiv:2205.08876 [gr-qc]].

\bibitem{Sheykhi:2010zz}
A.~Sheykhi,
Eur. Phys. J. C \textbf{69}, 265-269 (2010)
doi:10.1140/epjc/s10052-010-1372-9
[arXiv:1012.0383 [hep-th]].

\bibitem{Sheykhi:2021fwh}
A.~Sheykhi,
Phys. Rev. D \textbf{103}, no.12, 123503 (2021)
doi:10.1103/PhysRevD.103.123503
[arXiv:2102.06550 [gr-qc]].

\bibitem{Sheykhi:2023aqa}
A.~Sheykhi,
Phys. Lett. B \textbf{850}, 138495 (2024)
doi:10.1016/j.physletb.2024.138495
[arXiv:2302.13012 [gr-qc]].

\bibitem{Chatterjee:2020iuf}
A.~Chatterjee and A.~Ghosh,
Phys. Rev. Lett. \textbf{125}, no.4, 041302 (2020)
doi:10.1103/PhysRevLett.125.041302
[arXiv:2007.15401 [gr-qc]].

\bibitem{Padmanabhan2005}
T. Padmanabhan, 
``Gravity and the thermodynamics of horizons,'' 
\emph{Phys. Rept.} \textbf{406}, 49 (2005). 
doi:10.1016/j.physrep.2004.10.003



\bibitem{AkbarCai2007}
M. Akbar and R. G. Cai, 
``Thermodynamic behavior of field equations for $f(R)$ gravity,'' 
\emph{Phys. Lett. B} \textbf{648}, 243 (2007). 
doi:10.1016/j.physletb.2007.03.005

\bibitem{Hayward2006}
S. A. Hayward, 
``Formation and evaporation of non-singular black holes,'' 
\emph{Phys. Rev. Lett.} \textbf{96}, 031103 (2006). 
doi:10.1103/PhysRevLett.96.031103

\bibitem{Tsallis1988}
C. Tsallis, 
``Possible generalization of Boltzmann–Gibbs statistics,'' 
\emph{J. Stat. Phys.} \textbf{52}, 479 (1988). 
doi:10.1007/BF01016429



\bibitem{Lymperis2022}
A. Lymperis, 
``Cosmology through a Kaniadakis extended holographic dark energy model,'' 
\emph{Eur. Phys. J. C} \textbf{82}, 449 (2022). 
doi:10.1140/epjc/s10052-022-10443-4

\bibitem{Alruwaili:2025bwf} A. D. Alruwaili, Zoya Khan and Abdul Jawad, Nucl. Phys. B \textbf{1018}, 117032 (2025) doi: 10.1016/j.nuclphysb.2025.117032.

\bibitem{Jawad:2019vqa} A. Jawal, Z. Khan, S. Rani K. Bamba, Entropy \textbf{21}, 851 (2019) doi: 10.3390/e21090851.


\bibitem{KubiznakMann2012}
D. Kubizňák and R. B. Mann, 
``P–V criticality of charged AdS black holes,'' 
\emph{JHEP} \textbf{1207}, 033 (2012). 
doi:10.1007/JHEP07(2012)033

\bibitem{WeiLiu2020}
S. W. Wei and Y. X. Liu,  
\emph{Phys. Rev. D} \textbf{101}, 104018 (2020). 
doi:10.1103/PhysRevD.101.104018

\bibitem{jzhang} J. Zhang, Phys. Lett. B \textbf{668}, 353 (2028).

\bibitem{Ghosh:2012jf}
A.~Ghosh and P.~Mitra,
Phys. Lett. B \textbf{734}, 49-51 (2014)
doi:10.1016/j.physletb.2014.05.030
[arXiv:1206.3411 [gr-qc]].

\bibitem{Faraoni:2011hf}
V.~Faraoni,
Phys. Rev. D \textbf{84}, 024003 (2011)
doi:10.1103/PhysRevD.84.024003
[arXiv:1106.4427 [gr-qc]].

\bibitem{Faraoni:2015ula}
V.~Faraoni,
Lect. Notes Phys. \textbf{907}, pp.1-199 (2015)
2015,
ISBN 978-3-319-19239-0, 978-3-319-19240-6
doi:10.1007/978-3-319-19240-6

\bibitem{Okcu:2024llu}
{\"O}.~{\"O}kc{\"u} and E.~Aydiner,
Gen. Rel. Grav. \textbf{56}, no.7, 87 (2024)
doi:10.1007/s10714-024-03273-1
[arXiv:2407.14685 [gr-qc]].

\bibitem{Kubiznak:2012wp}
    D. Kubiznak and R. B Mann, JHEP \textbf{07}, 033 (2012) doi: 10.1007/JHEP07(2012)033 [arXiv:1205.0559 [hep-th]].

\bibitem{Helou:2015yqa}
A.~Helou,
[arXiv:1502.04235 [gr-qc]].

\bibitem{Helou:2015zma}
A.~Helou,
[arXiv:1505.07371 [gr-qc]].

\bibitem{Banks:2008ep}
T.~Banks,
J. Phys. A \textbf{42}, 304002 (2009)
doi:10.1088/1751-8113/42/30/304002
[arXiv:0809.3951 [hep-th]].

\bibitem{Scolnic:2021amr}
D.~Scolnic, D.~Brout, A.~Carr, A.~G.~Riess, T.~M.~Davis, A.~Dwomoh, D.~O.~Jones, N.~Ali, P.~Charvu and R.~Chen, \textit{et al.}
Astrophys. J. \textbf{938}, no.2, 113 (2022)
doi:10.3847/1538-4357/ac8b7a
[arXiv:2112.03863 [astro-ph.CO]].

\bibitem{Brout:2022vxf}
D.~Brout, D.~Scolnic, B.~Popovic, A.~G.~Riess, J.~Zuntz, R.~Kessler, A.~Carr, T.~M.~Davis, S.~Hinton and D.~Jones, \textit{et al.}
Astrophys. J. \textbf{938}, no.2, 110 (2022)
doi:10.3847/1538-4357/ac8e04
[arXiv:2202.04077 [astro-ph.CO]].

\bibitem{Mukherjee:2025xuy}
D.~Mukherjee, H.~S.~Sahota and S.~Gavas,
[arXiv:2509.04964 [gr-qc]].



\bibitem{Jimenez2003}
R.~Jimenez, L.~Verde, T.~Treu and D.~Stern,
Astrophys. J. \textbf{593}, 622-629 (2003)
doi:10.1086/376595
[arXiv:astro-ph/0302560 [astro-ph]].

\bibitem{Simon2005}
J.~Simon, L.~Verde and R.~Jimenez,
Phys. Rev. D \textbf{71}, 123001 (2005)
doi:10.1103/PhysRevD.71.123001
[arXiv:astro-ph/0412269 [astro-ph]].

\bibitem{Stern2010}
D.~Stern, R.~Jimenez, L.~Verde, M.~Kamionkowski and S.~A.~Stanford,
JCAP \textbf{02}, 008 (2010)
doi:10.1088/1475-7516/2010/02/008
[arXiv:0907.3149 [astro-ph.CO]].

\bibitem{Moresco2012}
M.~Moresco, A.~Cimatti, R.~Jimenez, L.~Pozzetti, G.~Zamorani, M.~Bolzonella, J.~Dunlop, F.~Lamareille, M.~Mignoli and H.~Pearce, \textit{et al.}
JCAP \textbf{08}, 006 (2012)
doi:10.1088/1475-7516/2012/08/006
[arXiv:1201.3609 [astro-ph.CO]].

\bibitem{Zhang2014}
C.~Zhang, H.~Zhang, S.~Yuan, T.~J.~Zhang and Y.~C.~Sun,
Res. Astron. Astrophys. \textbf{14}, no.10, 1221-1233 (2014)
doi:10.1088/1674-4527/14/10/002
[arXiv:1207.4541 [astro-ph.CO]].

\bibitem{Moresco2015}
M.~Moresco,
Mon. Not. Roy. Astron. Soc. \textbf{450}, no.1, L16-L20 (2015)
doi:10.1093/mnrasl/slv037
[arXiv:1503.01116 [astro-ph.CO]].

\bibitem{Moresco2016}
M.~Moresco, L.~Pozzetti, A.~Cimatti, R.~Jimenez, C.~Maraston, L.~Verde, D.~Thomas, A.~Citro, R.~Tojeiro and D.~Wilkinson,
JCAP \textbf{05}, 014 (2016)
doi:10.1088/1475-7516/2016/05/014
[arXiv:1601.01701 [astro-ph.CO]].


\bibitem{Ratsimbazafy2017}
A.~L.~Ratsimbazafy, S.~I.~Loubser, S.~M.~Crawford, C.~M.~Cress, B.~A.~Bassett, R.~C.~Nichol and P.~V{\"a}is{\"a}nen,
Mon. Not. Roy. Astron. Soc. \textbf{467}, no.3, 3239-3254 (2017)
doi:10.1093/mnras/stx301
[arXiv:1702.00418 [astro-ph.CO]].










\end{thebibliography}
\end{document}